# Electromagnetic enhancement of one-dimensional plasmonic hotspots along silver nanowire dimer examined by ultrafast surface enhanced fluorescence

Tamitake Itoh[1]*, Yuko S. Yamamoto[2]

[1]Health and Medical Research Institute, National Institute of Advanced Industrial Science and Technology (AIST), Takamatsu, Kagawa 761-0395, Japan

[2]School of Materials Science, Japan Advanced Institute of Science and Technology (JAIST), Nomi, Ishikawa 923-1292, Japan

*Corresponding author: tamitake-itou@aist.go.jp

**Abstract**

We investigated the spectral properties of electromagnetic (EM) enhancement of one-dimensional hotspots (1D HSs) generated between silver nanowire (NW) dimers. The EM enhancement spectra were directly derived by dividing the spectra of ultrafast surface-enhanced fluorescence (UFSEF) from single NW dimers with UFSEF obtained from large nanoparticle aggregates, which aggregate-by-aggregate variations in the UFSEF spectra were averaged out. Some NW dimers were found to exhibit EM enhancement spectra that deviated from the plasmon resonance Rayleigh scattering spectra, indicating that their EM enhancement was not generated by superradiant plasmons. These experimental results were examined by numerical calculation based on the EM mechanism by varying the morphology of the NW dimers. The calculations reproduced the spectral deviations as the NW diameter dependence of EM enhancement. Phase analysis of the enhanced EM near fields along the 1D HSs revealed that the dipole-quadrupole coupled plasmon, which is a subradiant mode, mainly generates EM enhancement for dimers with NW diameters larger than ~80 nm, which was consistent with scanning electron microscopic measurements.

**I. Introduction**

Effective cross-sections of optical responses for molecules inside the nanogaps or crevasses of nanoparticle (NP) aggregates are enhanced upon interaction with light tightly confined by plasmon resonance.[1] This phenomenon is called electromagnetic (EM) enhancement, which is expressed using the Purcell factor $F$, in addition to the radiative $\Delta\omega_R$ and nonradiative $\Delta\omega_{NR}$ decay rates of plasmon resonance as,[1]

$$F_R = \frac{F\Delta\omega_R}{\Delta\omega_R + \Delta\omega_{NR}}. \quad (1)$$

Here, $F$ is expressed as $F = \frac{Q(\lambda/n)^3}{4\pi^2 V_P}$, where $Q$ is the quality factor of the plasmon resonance, $\lambda$ and $n$ are the light wavelength and refractive index around the nanostructure, respectively, and $V_P$ is the mode volume of the plasmon. Equation (1) indicates that an extremely small $V_P$ in the nanogaps results in a large $F_R$. Particularly, spontaneous resonant Raman scattering comprising excitation and emission transitions undergoes twofold EM enhancement as $F_R(\lambda_{ex})F_R(\lambda_{em})$, where $\lambda_{ex}$ and $\lambda_{em}$ are the excitation and emission wavelengths, respectively.[2] The nanogaps inside NP aggregates reach a value of <$10^{10}$ for $F_R^2$,[3,4] enabling single molecule Raman spectroscopy.[5-9] Such nanogaps are referred to as "hotspots" (HSs). HSs have been applied to various surface-enhanced spectroscopic methods such as surface-enhanced (resonant) Raman scattering (SE(R)RS), absorption (SEA), fluorescence (SEF), and their nonlinear counterparts.[1] Furthermore, the EM coupling energy between plasmon resonance and molecular resonance at the HSs exceeds several hundred meVs,[10-14] resulting in the generation of new physical and chemical processes, for example, molecular optomechanics[15] and polariton chemistry.[16] The values of $V_P$ of HSs reach several tens of $nm^3$,[17,18] resulting in instability of enhanced optical signals.[19,20] To resolve the problem while maintaining a large $F_R$, several HSs have been theoretically and experimentally developed.[21,22] We have also developed one-dimensional HSs (1D HSs) where $V_P$ is extended by several $10^4$ times using the gaps or crevasses between silver nanowire (NW) dimers.[23,24] In such development, the evaluation of the spectral properties of $F_R$ is crucial for the clarification and optimization of these surface-enhanced spectroscopies and their related phenomena.[22]

Rayleigh scattering cross-section $\sigma_{sca}$ and extinction cross-section $\sigma_{ext}$ spectroscopy have been used to evaluate the $F_R$ of HSs.[1] In quantum optics, the interaction between a photon and a molecular exciton is expressed by the $\hat{\mathbf{A}} \cdot \hat{\mathbf{p}}$ and $\hat{\mathbf{A}}^2$ terms, where $\hat{\mathbf{A}}$ and $\hat{\mathbf{p}}$ are the operators of the vector potential and momentum, respectively.[25] The optical absorption (or emission) and Rayleigh scattering are expressed by the $\hat{\mathbf{A}} \cdot \hat{\mathbf{p}}$ and $\hat{\mathbf{A}}^2$ terms, respectively.[25] Fluorescence, Raman, and their nonlinear counterparts are described as two or more $\hat{\mathbf{A}} \cdot \hat{\mathbf{p}}$ terms.[26] Thus, the optical absorption cross-section $\sigma_{abs}$ spectroscopy of HSs should be use to evaluate $F_R$ instead of $\sigma_{sca}$ and $\sigma_{ext}$ spectroscopies. Indeed, the spectral relationship between $\sigma_{sca}$ (or $\sigma_{ext}$) and $F_R$ becomes unclear with increasing size or degree of asymmetry of the aggregates, because subradiant plasmon resonance, which does not explicitly appear in $\sigma_{sca}$, becomes dominant to generate $F_R$.[27,28] Thus, $\sigma_{abs}$ spectroscopy was developed using the relationship $\sigma_{abs} = \sigma_{ext} - \sigma_{sca}$.[29,30] However, this $\sigma_{abs}$ spectroscopy is applicable only for small NP aggregates. In the case of large NP aggregates, almost all incident photons become extinct, resulting in the saturation of $\sigma_{ext}$ and underestimation of $\sigma_{abs}$. Furthermore, $\sigma_{abs}$ spectroscopy provides absorption of whole NP aggregates, including those outside HSs, resulting in the $F_R$ of HSs becoming unclear in the $\sigma_{abs}$ spectra.[29]

In this study, we developed a spectroscopic technique to directly measure the $F_R$ of HSs using the ultrafast surface-enhanced fluorescence (UFSEF) of dye molecules inside HSs. UFSEF appears as the spectral background of SERRS when the SEF rate is faster than the vibrational decay rate of the excited electronic states.[31,32] We applied this spectroscopy to measure the $F_R$ of the 1D HSs of the NW dimers, to which the $\sigma_{abs}$ spectroscopy is not applicable. The $F_R$ spectra were derived by dividing the spectra of UFSEF from the 1D HSs of single NW dimers by the UFSEF of large NP aggregate, where aggregate-by-aggregate spectral variations in $F_R$ were averaged out. The derived $F_R$ spectra frequently exhibit blue shifts from the $\sigma_{sca}$ spectra, suggesting that such $F_R$ is not generated by superradiant plasmon but by subradiant plasmon. These results were examined using numerical calculation based on the EM mechanism by changing the NW diameter of the symmetric NW dimers and the degree of asymmetry of NW dimers. Calculations performed by changing the diameter reproduced the blueshifts in $F_R$ well. The phase retardation of enhanced electric near-fields against incident fields was -180°, which indicates that the dipole-quadrupole coupled plasmon resonance,[27] which is subradiant, mainly generates $F_R$ for dimers where the NW diameters are larger than 80 nm. These results are consistent with the scanning electron microscopic measurements.

**II. Theoretical background**

Here, we briefly explain the mechanism of UFSEF. The total radiative decay rate of a dye molecule in a free space is $\Gamma_{r0}$ (= $\int \gamma_{r0}(\lambda_{em})d\lambda_{em}$), where $\gamma_{r0}(\lambda_{em})d\lambda_{em}$ is the radiative decay rate at $\lambda_{em}$.[31,32] The total radiative decay rates of a molecule inside a HS enhanced by $F_R$ are $\Gamma_{Rad}$ (= $\int \gamma_{r0}(\lambda_{em})F_R(\lambda_{em})d\lambda_{em}$). The total internal relaxation rate of a molecule $\Gamma_{int} \sim 10^{12}$ s$^{-1}$ is generally significantly larger than $\Gamma_{r0} \sim 10^{8-9}$ s$^{-1}$ for fluorescent dye molecules.[31,32] Thus, the fluorescent molecule emits a photon from the bottom of the vibrational state in the electronic excited state (S$_1$), as shown in Fig. 1(a1).[32] Thus, the fluorescence spectrum does not have excitation laser wavelength dependence, as shown in Fig. 1(a2). However, $F_R \sim 10^5$ inside the HSs yields $\Gamma_{int}$ ($10^{12}$ s$^{-1}$) < $\Gamma_{Rad}$ ($10^5 \times 10^9$ s$^{-1}$), indicating that SEF has a component of photons emitted from the vibrational excited state in $S_1$ before relaxing to the bottom of $S_1$, as shown in Fig. 1(b1).[32] Such emission is called UFSEF, and appears as the background of the SERRS spectrum. The UFSEF spectra exhibit blue shifts in the highest energy from conventional fluorescence and becomes dependent on the excitation laser wavelength as shown in Fig. 1(b2).[32]

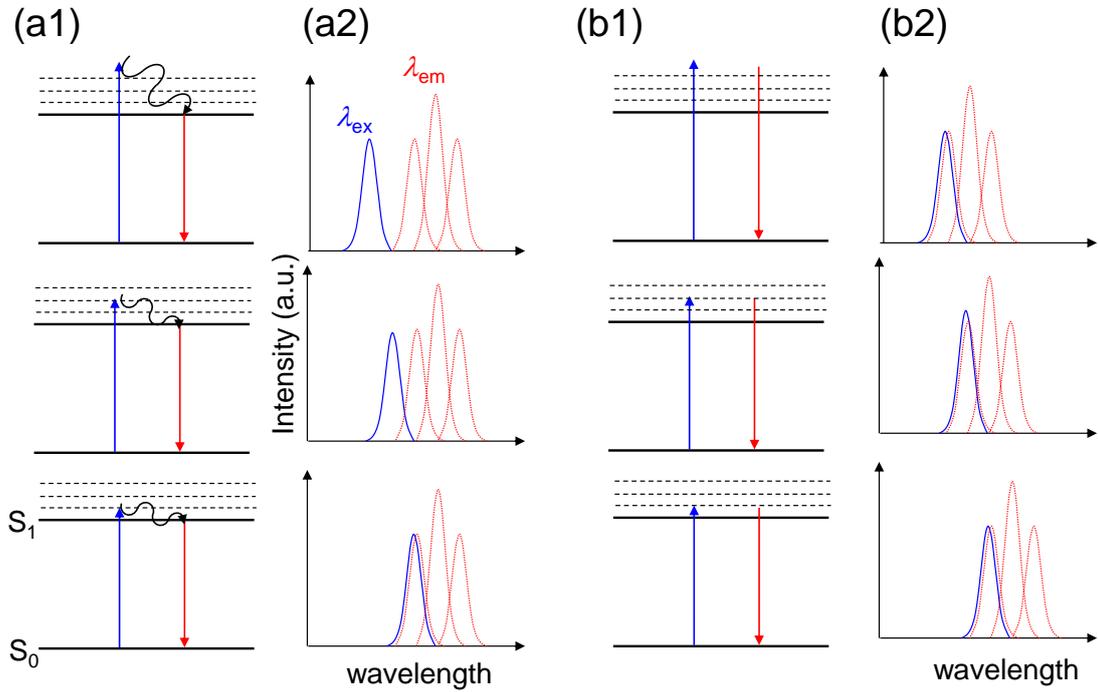

FIG. 1. (a1) Energy-level diagrams for excitation transition by laser light (blue arrows) and conventional SEF transition ($\Gamma_{int} \gg \Gamma_{Rad}$) (red arrows). (a2) Envelopes of a blue

and three red curves correspond to excitation and fluorescent spectra, respectively. Vibrational relaxations are indicated by wavy arrows. (b1) Energy-level diagrams for excitation transition by laser light (blue arrows) and UFSEF transition ($\Gamma_{int} < \Gamma_{Rad}$) (red arrows). (b2) Envelopes of a blue and three red curves correspond to UFSEF spectra. $S_0$ and $S_1$ indicate the ground and lowest excited electronic states. Dashed lines indicate vibrational excited states.

The relationships between the $\hat{\mathbf{A}} \cdot \hat{\mathbf{p}}$ and $\hat{\mathbf{A}}^2$ terms and molecular optical processes (i.e., absorption, Rayleigh scattering, fluorescence, and Raman scattering) are explained to show the importance of the $\hat{\mathbf{A}} \cdot \hat{\mathbf{p}}$ term for fluorescence and Raman or UFSEF and SE(R)RS.[26] Here, $\hat{H}_{int}$, which indicates the Hamiltonian for EM interactions between the electrons and the EM field, is described as follows:

$$\hat{H}_{int} = \sum_{i=1}^{N} -\frac{e_i}{m_i}\hat{\mathbf{A}}(\mathbf{r}_i) \cdot \hat{\mathbf{p}}_i + \sum_{i=1}^{N} \frac{e_i^2}{2m_i}\hat{\mathbf{A}}^2(\mathbf{r}_i), \quad (2)$$

where $e_i$, $m_i$, and $\mathbf{r}_i$ are the charge, effective mass, and position of the $i$-th electron, respectively.[25] The first and second terms on the right side of Eq. (2) are called the $\hat{\mathbf{A}} \cdot \hat{\mathbf{p}}$ term ($\hat{H}_{int}(AP)$) and $\hat{\mathbf{A}}^2$ term ($\hat{H}_{int}(A^2)$), respectively. The time-independent Schrödinger equation for the stationary states of a system without EM interactions is

$$\left(\hat{H}_e + \hat{H}_{ph}\right)\Psi_0^n(\mathbf{r}) = E_0^n \Psi_0^n(\mathbf{r}), \quad n = 1, 2, 3, \cdots, \quad (3)$$

where $\hat{H}_e$ and $\hat{H}_{ph}$ are Hamiltonians for the electrons and the EM field, respectively, and $\Psi_0^n(\mathbf{r})$ and $E_0^n$ represent the wave function and energy of the system in the *n*-th state (superscript) for the unperturbed system (subscript *0*), respectively.[25] Assuming that the system initially in $\Psi_0^i$ at $t = 0$ starts to be perturbed by $\hat{H}_{int}$, the probability $|b_f(t)|^2$ for the system finally to be $\Psi_0^f$ can be derived using the first- and second-order perturbation terms as

$$|b_f(t)|^2 \propto \left| \langle \Psi_0^f | \hat{H}_{int} | \Psi_0^i \rangle + \sum_{n \neq i} \langle \Psi_0^f | \hat{H}_{int} | \Psi_0^n \rangle \langle \Psi_0^n | \hat{H}_{int} | \Psi_0^i \rangle \right|^2, \quad (4)$$

where $\langle \Psi_0^f | \hat{H}_{int} | \Psi_0^i \rangle$ and $\sum_{n \neq i} \langle \Psi_0^f | \hat{H}_{int} | \Psi_0^n \rangle \langle \Psi_0^n | \hat{H}_{int} | \Psi_0^i \rangle$ indicate the transitions related to the first- and second-order perturbation terms, respectively, and the subscripts *f* and *n* denote the final and intermediate states, respectively.[26]

The matrix element $\langle \Psi_0^f | \hat{H}_{int} | \Psi_0^i \rangle$ in Eq. (4) can be separated into $\langle \Psi_0^f | \hat{H}_{int}(AP) | \Psi_0^i \rangle$ and $\langle \Psi_0^f | \hat{H}_{int}(A^2) | \Psi_0^i \rangle$ using Eq. (2). $\langle \Psi_0^f | \hat{H}_{int}(AP) | \Psi_0^i \rangle$ corresponds to one-photon absorption or one-photon emission. It is well-known as the Fermi's golden rule, and is given by

$$\langle \Psi_0^f | \hat{H}_{int}(AP) | \Psi_0^i \rangle \propto \sum_{i=1}^{N} \frac{\mathbf{i}}{\hbar}(\varepsilon_f - \varepsilon_i)\langle f | e_i \mathbf{r}_i | i \rangle, \quad (5)$$

where $|i\rangle$ and $|f\rangle$ are the eigenfunctions of the initial and final states of the electron system, respectively, and $\varepsilon_i$ and $\varepsilon_f$ are their respective energies. $\langle f|e_i\mathbf{r}_i|i\rangle \neq 0$ can be nonzero for $|i\rangle \neq |f\rangle$.[26] The $\hat{\mathbf{A}}^2$ terms $\langle\Psi_0^f|\hat{H}_{int}(A^2)|\Psi_0^i\rangle$ can be described as

$$\langle\Psi_0^f|\hat{H}_{int}(A^2)|\Psi_0^i\rangle \propto \sum_{i=1}^{N}\frac{\mathbf{i}}{\hbar}\langle f||i\rangle. \quad (6)$$

Here, $\langle f||i\rangle$ can be nonzero for $|i\rangle = |f\rangle$, meaning that the two-photon process in Eq. (6) corresponds to Rayleigh scattering.[26]

The transitions are evaluated using the second-order perturbation term $\sum_{n\neq i}\langle\Psi_0^f|\hat{H}_{int}|\Psi_0^n\rangle\langle\Psi_0^n|\hat{H}_{int}|\Psi_0^i\rangle$ in Eq. (4). The matrix elements corresponding to the Rayleigh and Raman scattering should include $\hat{H}_{int}(AP)$ as a nonzero term. Thus, this term can be rewritten as:

$$\sum_{n\neq i}\langle\Psi_0^f|\hat{H}_{int}(AP)|\Psi_0^n\rangle\langle\Psi_0^n|\hat{H}_{int}(AP)|\Psi_0^i\rangle. \quad (7)$$

The matrix elements in Eq. (7) mainly contribute to Raman scattering for $|i\rangle \neq |f\rangle$.[26]

In summary, absorption (or emission) and Rayleigh scattering are generated by $\langle\Psi_0^f|\hat{H}_{int}(AP)|\Psi_0^i\rangle$ and $\langle\Psi_0^f|\hat{H}_{int}(A^2)|\Psi_0^i\rangle$, respectively, as shown in Figs. 2(a) and 2(b). Fluorescence involves a two-step process of two terms of $\langle\Psi_0^f|\hat{H}_{int}(AP)|\Psi_0^i\rangle$, as shown in Fig. 2(c). Raman scattering occurs via $\sum_{n\neq i}\langle\Psi_0^f|\hat{H}_{int}(AP)|\Psi_0^n\rangle\langle\Psi_0^n|\hat{H}_{int}(AP)|\Psi_0^i\rangle$, as shown in Fig. 2(d). Regarding the contribution of $\hat{H}_{int}(AP)$ to fluorescence and Raman scattering, the spectral shapes of $F_R$ in Eq. (1) are correlated to UFSEF and SE(R)RS.

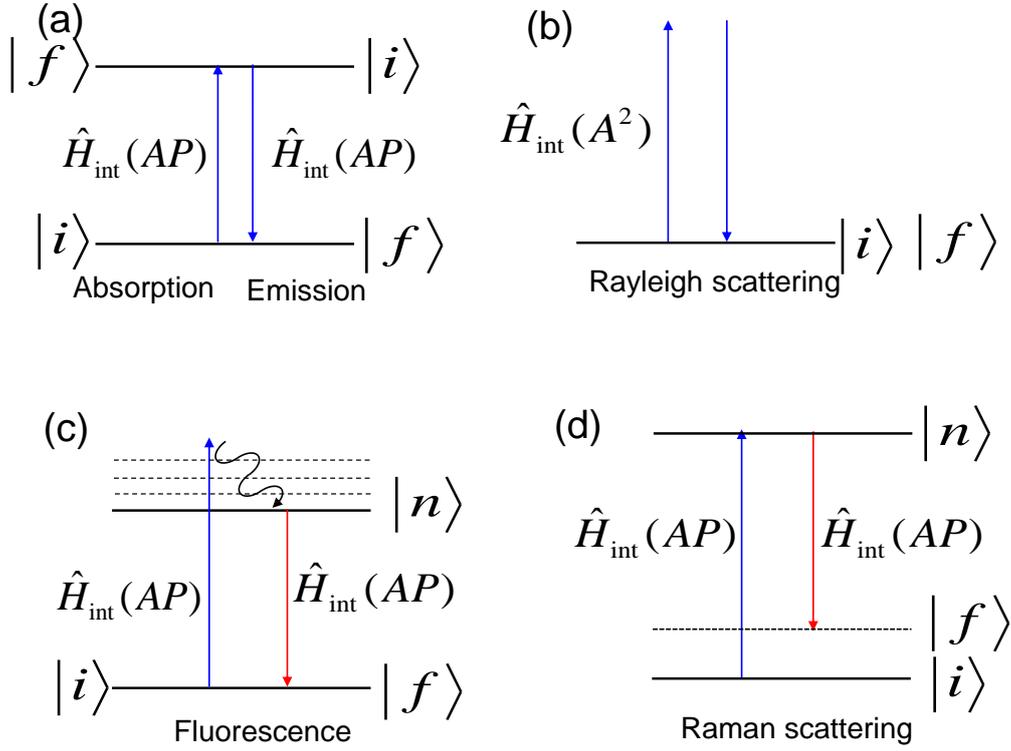

FIG. 2. (a) Energy-level diagrams of absorption and emission transitions from initial $|i\rangle$ to final state $|f\rangle$, respectively, triggered by $\hat{H}_{int}(AP)$. (b) Energy-level diagram of Rayleigh scattering transition from $|i\rangle$ to $|f\rangle$ ($|i\rangle = |f\rangle$) triggered by $\hat{H}_{int}(A^2)$. (c) Energy-level diagram of fluorescence process composed of absorption, vibrational relaxation (wavy arrow), and emission transitions from $|i\rangle$ to $|f\rangle$ via $|n\rangle$ by $\hat{H}_{int}(AP)$. Dashed lines indicate vibrational excited states. (d) Energy-level diagram of Raman process composed of absorption and emission transitions from $|i\rangle$ to $|f\rangle$ via $|n\rangle$ by $\hat{H}_{int}(AP)$. Dashed line indicates vibrational excited state $|f\rangle$.

**III. Materials and experimental methods**

The preparation methods of silver NW and NP samples are as described previously.[33,34] The average NW diameter and length are approximately 60 nm and 10 μm, respectively. The NW suspension was dropped and dried on a glass plate with a rhodamine 6G (R6G) methanol solution (~5.0 × $10^{-6}$ M). The effective concentration of the dye on the NWs was reduced by photobleaching most of the dye molecules adsorbed on the NW and glass surfaces via irradiation with a green laser beam. The reduction in concentration was confirmed by blinking and quenching of the UFSEF and fluorescence from both the NWs and glass surface, respectively.[23] The Rayleigh scattering spectra of the NW edges clearly exhibited oscillating structures due to the Fabry–Perot type resonance of the longitudinal surface plasmon mode, indicating that the NWs can be treated as single-crystalline rather than polycrystalline.[23]

The dark-field images were measured by focusing the white light of a 50-W halogen lamp through a dark-field condenser (numerical aperture (NA) 0.92) to collect Rayleigh scattering light from single NW dimers. Superradiant plasmon resonance maxima of the NW dimers were observed in the Rayleigh scattering spectra. Figure 3(a) shows the measurement of a UFSEF image to obtain an entire 1D HS image. Unpolarized excitation laser beams with wavelengths 457, 473, 532, 561, and 633 nm were loosely focused using an objective lens (NA 0.2). The green closed circle in Fig.

3(b) indicates the focusing spot area (~200 × 400 $\mu m^2$) of the laser beam on a schematic of the 1D HS. A suitable laser wavelength was selected to evaluate the $F_R$ spectra using UFSEF spectra. A 100× objective lens, as shown in Fig. 3(a), was used to collect both Rayleigh scattering and UFSEF light from the NW dimer. Figures 3(c1) and 3(d1) show the dark-field images, and Figs. 3(c2)–3(c6) and 3(d2)–3(d6) show the UFSEF images of single NW dimer and a large NP aggregate at five excitation laser wavelengths, respectively. The Rayleigh scattering light and UFSEF light from an identical position are sent to a polychromator for spectroscopy through a pinhole to selectively measure the position at which the area is approximately 1.8×1.8 $\mu m^2$, as indicated in Fig. 3(b) by the dashed black circle.

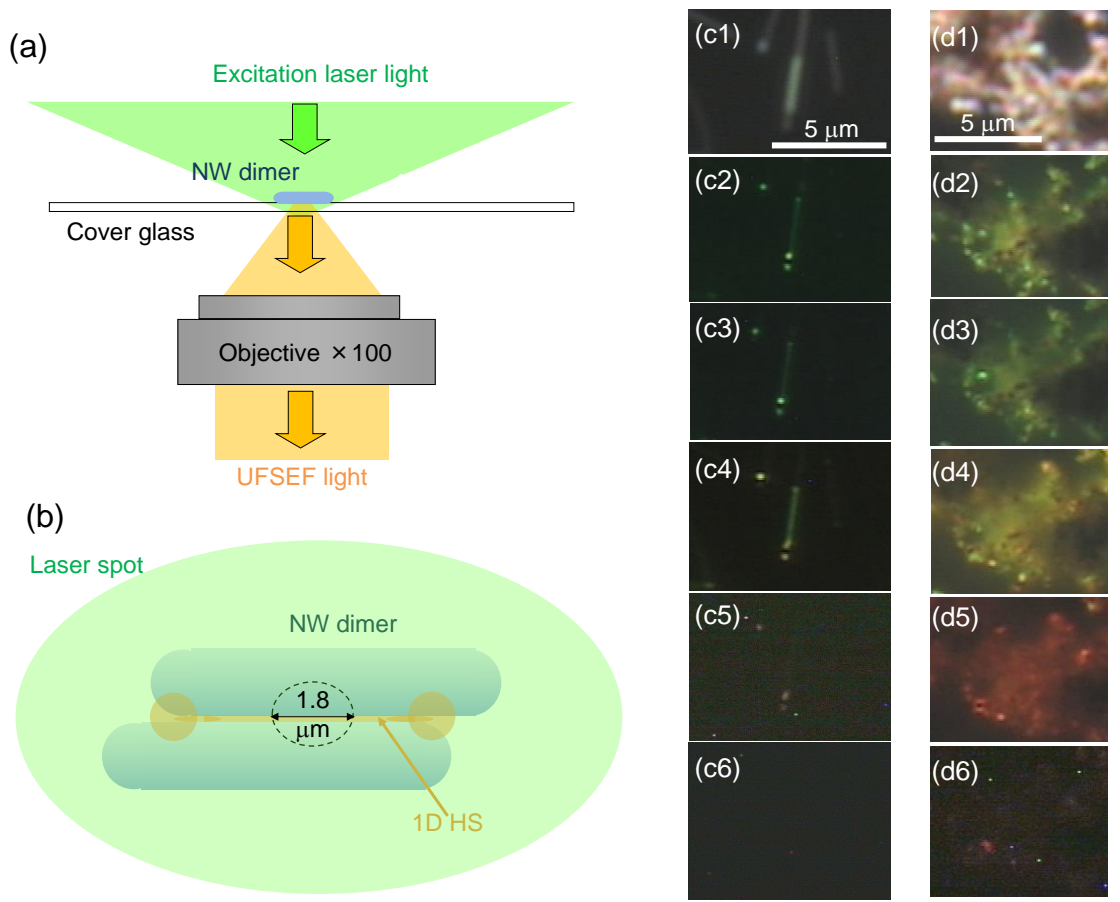

FIG. 3. (a) Experimental setup of laser beam excitation for NW dimer. NW dimers placed on the cover glass are excited from above and forward UFSEF light is detected. (b) Schematic of the UFSEF image of entire 1D HS illustrated with orange closed circles and a line. The large green spot indicates a focusing laser spot of excitation light. The dashed black open circle is detection area. (c1)–(c6) and (d1)–(d6) dark-field image (c1, d1), UFSEF images with the five excitation laser wavelengths 457 (c2, d2), 473 (c3, d3), 532 (c4, d4), 561 (c5, c5), 633 (c6, c6) nm, respectively of single NW dimer and large NP aggregates. Scale bars in (c1)–(c6) and (d1)–(d6) are 5.0 μm.

**IV. Results and discussion**

Figures 4(a)–4(c) show the scanning electron microscopy (SEM, JSM-6700F JEOL) image of the NWs on an ITO glass surface, dark-field image, and UFSEF image of an identical area. The blue green color of NW dimers in Fig. 4(b) indicates that the plasmons of two NWs are coupled and their coupled resonance redshifts to the visible region.[23] The UFSEF activity of NW dimers in Fig. 4(c) reveals that the coupled plasmon resonance generates $F_R$ along 1D HSs.[23] The UFSEF light propagates through 1D HSs, indicating both localized and propagating plasmons are involved in $F_R$.[24] Figure 4(d) shows the distribution of NW diameters estimated by the SEM images. Almost all NW diameters were distributed in the range of 40–90 nm. Figure 4(e) shows the degree of asymmetry of the NW dimers. The NW diameter ratios were within 2.0. These ratios are consistent with the distribution of the NW diameters shown in Fig. 4(d).

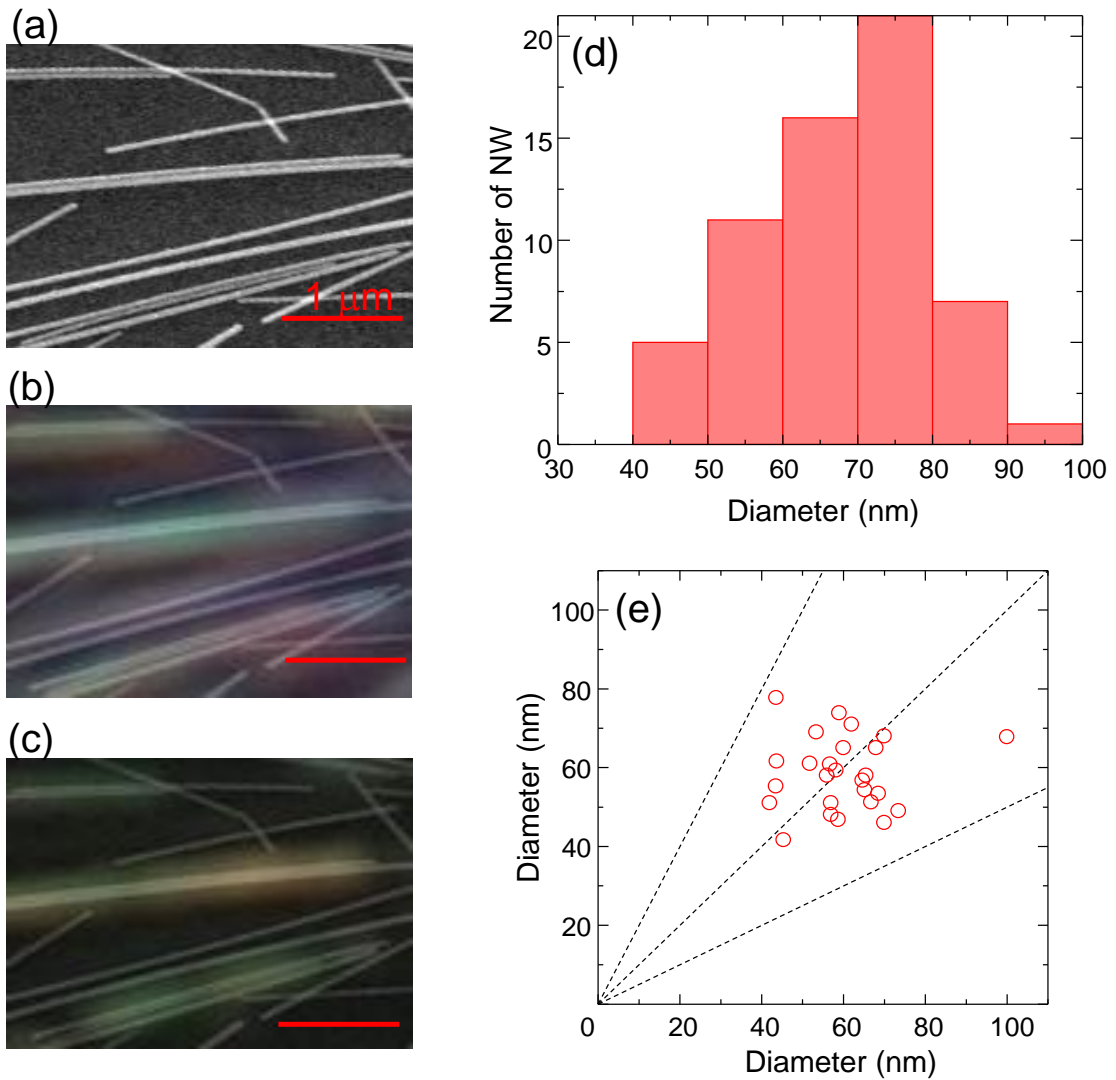

FIG. 4. (a) SEM image of NWs and NW dimers. (b) Superposition of the SEM and dark-field images. (c) Superposition of the SEM and UFSEF images. Scale bars in (a)–(c) are 1.0 μm. (d) Distribution of NW diameters estimated by SEM images. (e) Distribution of NW diameters of NW dimers estimated by SEM images. The three dashed lines indicate the diameter ratios between two NWs 0.5 (lower), 1.0 (middle), 2.0 (upper), respectively.

The UFSEF process comprises excitation and emission transitions, as shown in Fig. 2(c), expressed by two $\hat{\mathbf{A}} \cdot \hat{\mathbf{p}}$ terms.[26] Thus, the EM enhancement factor of the UFSEF is described as the product of the excitation enhancement factor $F_R(\lambda_{ex})$ and emission enhancement factor $F_R(\lambda_{em})$ using Eq. (1).[1] Thus, the cross-section of UFSEF $\sigma_{UFSEF}(\lambda_{ex}, \lambda_{em})$ is expressed as

$$\sigma_{UFSEF}(\lambda_{ex}, \lambda_{em}) = F_R(\lambda_{ex}) F_R(\lambda_{em}) \sigma_{UFflo}(\lambda_{ex}, \lambda_{em}), \quad (8)$$

where $\sigma_{UFflo}(\lambda_{ex}, \lambda_{em})$ is the cross-section of the ultrafast fluorescence without EM enhancement. Equation (8) indicates that the UFSEF spectrum is modulated by $F_R(\lambda_{em})$, which can originate from both superradiant and subradiant plasmon resonances. The purpose of this study was to experimentally extract $F_R$ spectra from this spectral modulation. Extraction was performed by dividing the UFSEF spectra from the 1D HSs of single NW dimers (Figs. 3(c2)–3(c6)) by the UFSEF of large NP aggregate (Figs. 3(d2)–3(d6)). Such NP aggregate contain a large number of HSs which $F_R(\lambda_{em})$ values exhibit different spectral shapes. Thus, the $F_R(\lambda_{em})$ of the large NP aggregate is averaged, and becomes constant.[19,32] Thus, its UFSEF spectrum can be expressed as

$$\propto F_R(\lambda_{ex}) \sigma_{UFflo}(\lambda_{ex}, \lambda_{em}). \quad (9)$$

Thus, $F_R$ spectrum of a single NW dimer can be experimentally derived by dividing Eq.

(8) by Eq. (9) as

$$F_R(\lambda_{em}) = \frac{F_R(\lambda_{ex})F_R(\lambda_{em})\sigma_{UFflo}(\lambda_{ex},\lambda_{em})}{F_R(\lambda_{ex})\sigma_{UFflo}(\lambda_{ex},\lambda_{em})}. \quad (10)$$

The UFSEF spectra of single-NW dimers were measured as per Eq. (8). Figure 5(a1) shows the $\sigma_{sca}$ spectrum of a single NW dimer. The maximum of $\sigma_{sca}$ at approximately 540 nm corresponds to the superradiant plasmon resonance. If the $F_R$ for UFSEF is generated by subradiant plasmon resonance, the maximum of the $F_R$ spectrum is not observed in the $\sigma_{sca}$ spectrum, but is observed in the UFSEF spectrum. This is because the UFSEF spectrum is modulated by the $F_R$ generated by both superradiant and subradiant resonances. Therefore, the UFSEF spectra of a single NW dimer were measured to determine the $F_R$ maximum using multiple laser excitations. Figures 5(a2)–5(a6) show the $\lambda_{ex}$ dependence of the UFSEF spectra for an identical NW dimer. The UFSEF spectral intensities were normalized to the excitation laser power. The normalized UFSEF intensity was the highest at $\lambda_{ex}$ of 532 nm, which is close to the maximum of the absorption spectrum of the R6G $\propto \sigma_{UFflo}(\lambda_{ex})$, indicating the resonance effect of the R6G molecules. However, the entire structures of the UFSEF spectra are observed in Figs. 5(a2) and 5(a3), indicating that the spectral maximum of the $F_R$ is located around these spectral regions.

The UFSEF spectra of large NP aggregates were measured as per Eq. (9).

Figure 5(b1) shows the $\sigma_{sca}$ spectrum of the aggregate. The spectrum appears flat, indicating that the $F_R$ spectra are averaged out and becomes also flat. Thus, one can obtain the UFSEF spectra corresponding to Eq. (9). Figures 5(b2)–5(b6) show the $\lambda_{ex}$ dependence of the UFSEF spectra of the aggregate. The UFSEF spectral intensities were normalized to the excitation laser power. The normalized UFSEF intensity reaches the maximum at $\lambda_{ex}$ 532 nm, which is also close to the absorption maximum of R6G, as shown in like Fig. 5(a4). It should be noted that the spectral properties in Figs. 5(b2)–5(b6) are different from those in Figs. 5(a2)–5(a6). For example, the UFSEF spectra in Figs. 5(b2) and 5(b3) appear significantly broader than those in Figs. 5(a2) and 5(a3). The UFSEF spectra in Figs. 5(a5) and 5(a6) show almost noise levels, but those in Figs. 5(b5) and 5(b6) clearly show the spectral structures. These different properties of the UFSEF spectra of the single NW dimer and large NP aggregate may be due to their $F_R$ spectra.

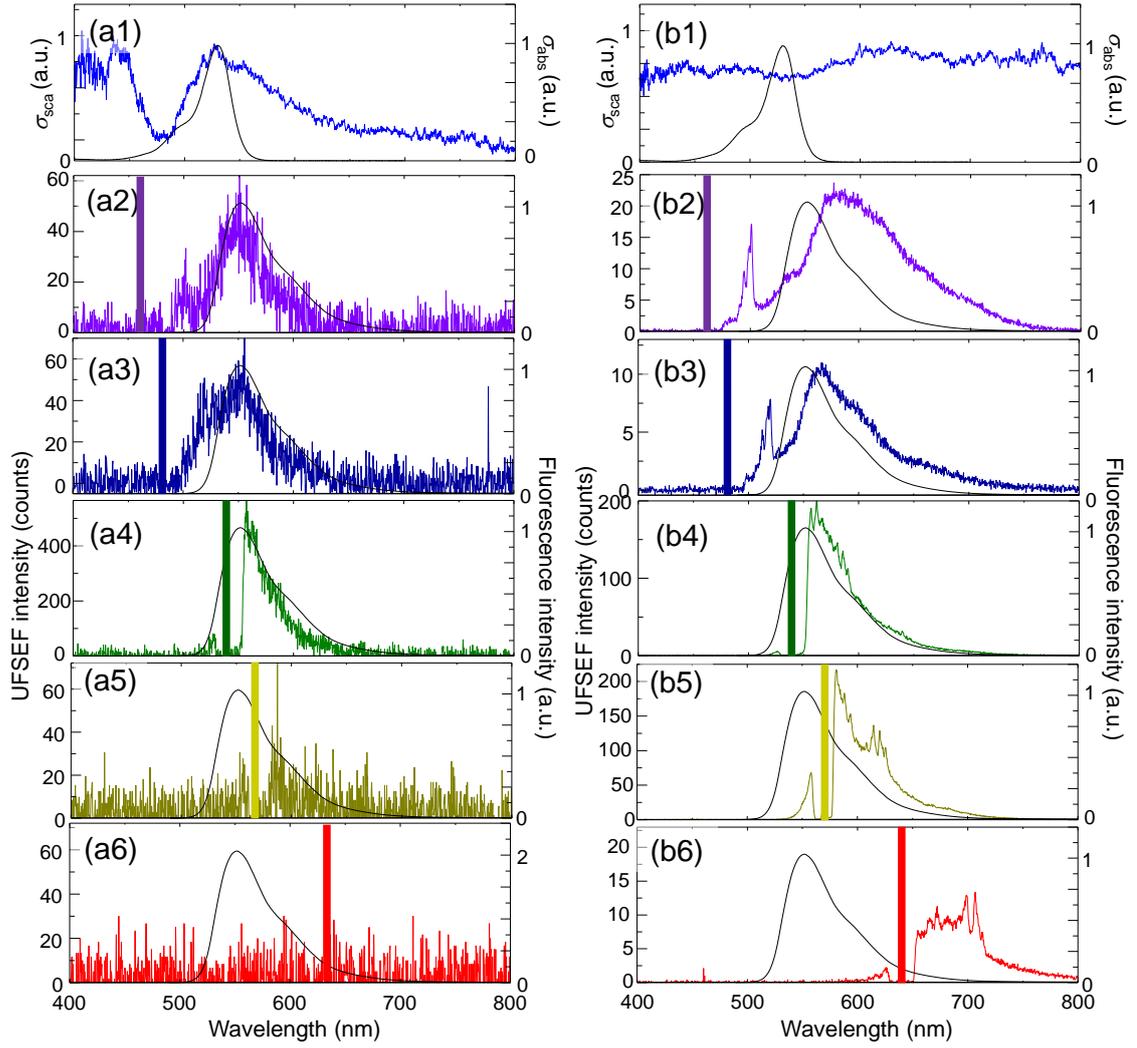

FIG. 5. (a1)–(a6) Spectra of $\sigma_{sca}$ (a1) (blue curve), and UFSEF with the five excitation laser wavelengths 457 (a2) (purple curve), 473 (a3) (dark blue curve), 532 (a4) (green curve), 561 (a5) (dark yellow curve), 633 (a6) (red curve) nm, respectively, for single NW dimer. (b1)–(b6) Spectra of $\sigma_{sca}$ (a1), and UFSEF with the five excitation laser wavelengths 457 (a2) (purple curve), 473 (a3) (dark blue curve), 532 (a4) (green curve), 561 (a5), 633 (a6) (red curve) nm, respectively, for a large NP aggregate. Black curves in (a1) and (b1) are absorption spectra of free R6G molecules. Black curves in (a2)–(a6)

and (b2)–(b6) are fluorescence spectra of free R6G molecules. Vertical lines curves in (a2)–(a6) and (b2)–(b6) indicate excitation laser wavelengths. Spike lines in (b2)–(b6) are SERRS spectra.

The procedure to experimentally obtain the $F_R$ spectra are illustrated in Fig. 6, using Eq. (10). Figures 6(a) and 6(b) show the UFSEF spectra of a single NW dimer and a large NP aggregate, respectively. Figures 6(a) and 6(b) correspond to the UFSEF spectra in Eqs. (8) and (9), respectively. According to Eq. (10), the $F_R$ spectrum in Fig. 6(c) was derived by dividing the UFSEF spectrum in Fig. 6(a) by that in Fig. 6(b). The contributions of superradiant and subradiant plasmon resonances to $F_R$ are evaluated by comparing the derived $F_R$ with the $\sigma_{sca}$ spectra of a single NW dimer. The contribution of the subradiant resonance appears as a deviation between the $F_R$ and $\sigma_{sca}$ spectra. We define the maximum wavelengths of the $\sigma_{sca}$ and $F_R$ spectra as $\lambda_{sca}$ and $\lambda_F$, respectively, as shown in Fig. 6(c).

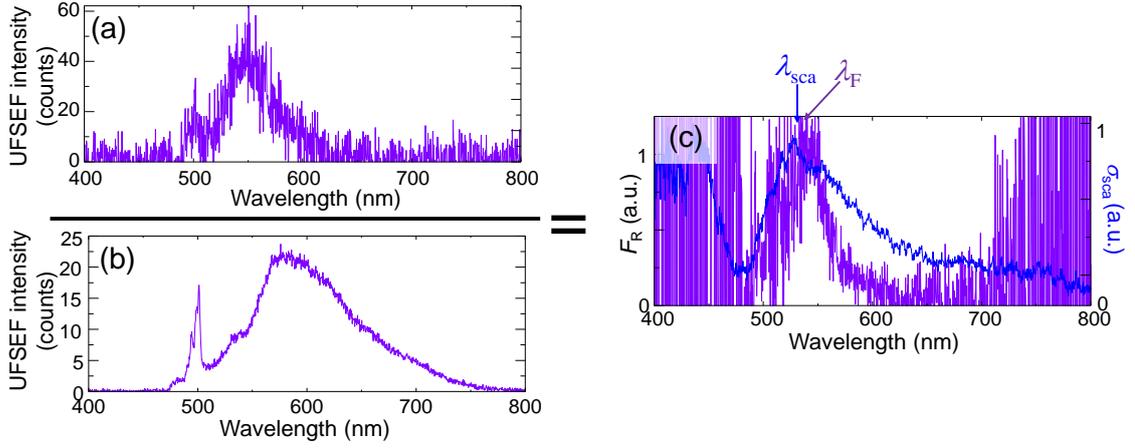

FIG. 6. Schematic for derivation of the $F_R$ spectrum based on Eqs. (8)–(10). (a) and (b) Spectra of UFSEF of single NW dimer corresponding to Eq. (8) and a large NP aggregate corresponding to Eq. (9), respectively. (c) Spectra of $F_R$ (purple curve) and $\sigma_{sca}$ (blue curve). The position of the maxima of $F_R$ and $\sigma_{sca}$ at $\lambda_F$ and $\lambda_{sca}$ are indicated in the panel, respectively. The spectral regions in < 480 and > 700 nm were noise owing to weak signals of UFSEF and insertion of notch filter.

The $F_R$ spectra were derived by dividing the UFSEF spectra of single NW dimers as shown in Figs. 5(a2)–5(a6) by the UFSEF spectra of the large NP aggregate as shown in Figs. 5(b2)–5(b6). Figures 7(a1)–7(a4) exhibit the obtained $\lambda_{ex}$ dependence of the $F_R$ spectra for a single NW dimer with $\sigma_{sca}$. With decreasing $\lambda_{ex}$, the entire envelope of $F_R$ spectrum appeared in the spectral region where the UFSEF had sufficient intensity. The positions of $\lambda_F$ appear near those of $\lambda_{sca}$ in Figs. 7(a1) and 7(a2).

The results $\lambda_F \sim \lambda_{sca}$ indicate that $F_R$ of this NW dimer was generated by superradiant plasmon resonance. However, the spectral slopes of $F_R$ around 550–700 nm are much steeper than those of $\sigma_{sca}$. We examined the $\lambda_{ex}$ dependence of $F_R$ for several single NW dimers and found that the spectral maxima in $F_R$ were always observed when $\lambda_{ex} = 457$ nm. Therefore, we measured the $F_R$ spectra of several single-NW dimers using this $\lambda_{ex}$, as shown in Figs 7(b1)–7(b5). Figures 7(b1) and 7(b2) show that $\lambda_F$ appears around $\lambda_{sca}$ for NW dimers for which $\lambda_{sca}$ is less than 550 nm. However, $\lambda_F$ remained at approximately 550 nm even when $\lambda_{sca}$ exceeded 550 nm, as shown in Figs. 7(b4) and 7(b5). This deviation between $\lambda_F$ and $\lambda_{sca}$ indicates the contribution of the subradiant plasmon resonance to $F_R$.

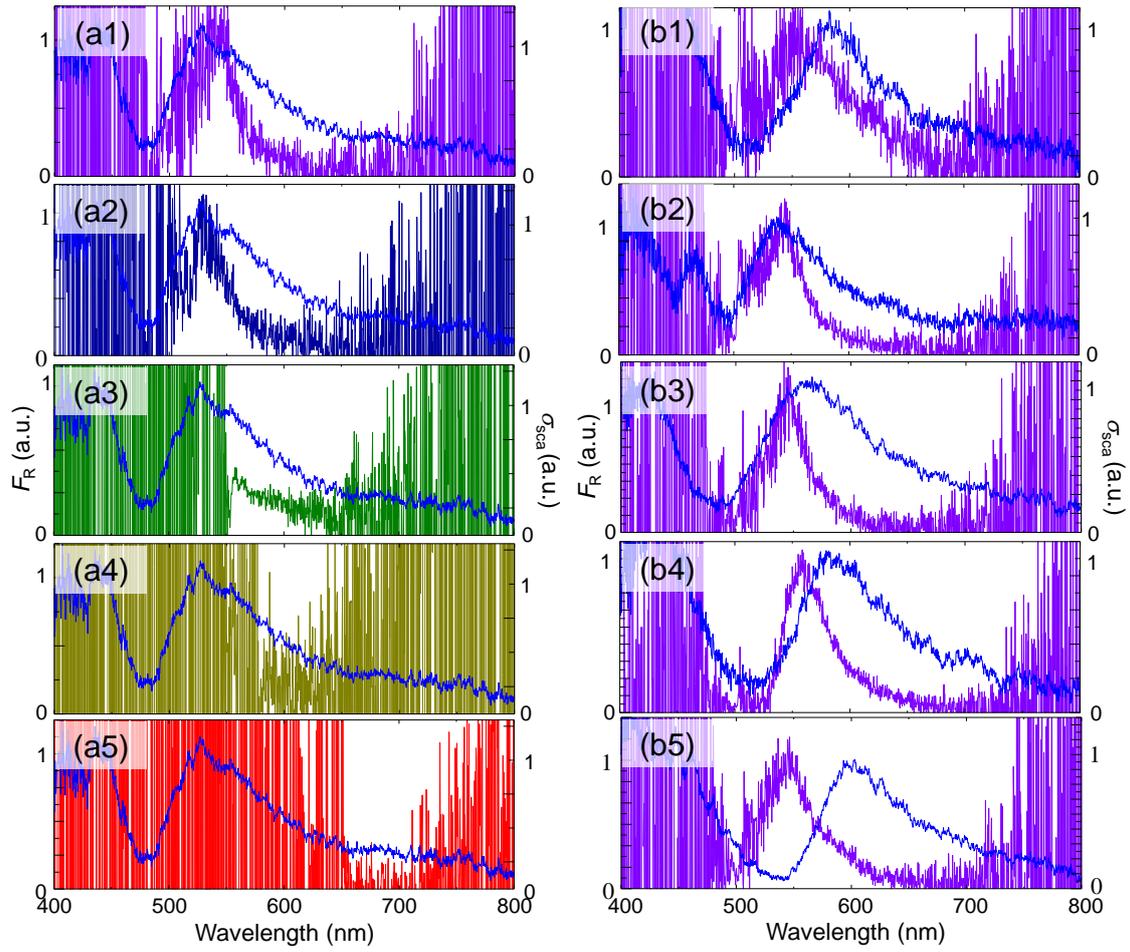

FIG. 7. (a1)–(a5) Spectra of $\sigma_{sca}$ (blue curve) and $F_R$ derived using UFSEF spectra of single NW dimer in FIG. 5. (a2)–(a6) and those of a large NP aggregate in FIG. 5. (b2)–(b6), respectively, based on Eq. (10) with the five excitation laser wavelengths 457 (a1) (purple curve), 473 (a2) (dark blue curve), 532 (a3) (green curve), 561 (a4) (dark yellow curve), 633 (a5) (red curve) nm, respectively. (b1)–(b5) Spectra of $\sigma_{sca}$ (blue curve) and $F_R(\lambda_{em})$ (purple curves) derived using UFSEF of five single NW dimers and a large NP aggregate in FIG. 5. (b2) based on Eq. (10) with the excitation laser wavelength 457 nm.

The relationship between $\lambda_F$ and $\lambda_{sca}$ was investigated to confirm the contribution of superradiant and subradiant plasmon resonances to $F_R$. Figure 8 shows this relationship using 23 NW dimers with $\lambda_{ex}$ of 457 nm. Two trends are observed, as indicated by the dashed lines. The first trend is that the $\lambda_F$ values are almost the same as the $\lambda_{sca}$ values for wavelengths below 550 nm. This trend indicates that superradiant plasmon resonance mainly generates $F_R$. The second trend is that the values of $\lambda_F$ remain constant at approximately 550 nm, even with an increase in the values of $\lambda_{sca}$ for wavelengths greater than 550 nm. This trend suggests the contribution of subradiant plasmon resonance to $F_R$.

The two trends observed in Fig. 8 indicate that the $F_R$ spectra were generated by superradiant plasmon resonance for $\lambda_{sca} < 550$ nm and subradiant plasmon resonance for $\lambda_{sca} > 550$ nm. If this indication is correct, these two trends can be reproduced using EM calculations as a spectral correlation or as an uncorrelation between $\sigma_{sca}$ and $F_R$. Thus, these trends were examined using the finite-difference time-domain (FDTD) method (EEM-FDM Version 5.1, EEM Co., Ltd., Japan). The complex refractive indices of the NWs were adopted from Ref. 35. The effective refractive index of the surrounding medium was set to 1.25 to ensure consistency between the calculated and experimental $\sigma_{sca}$ spectra of gold NPs.[26-29] The experimentally obtained spectral relationships between $\sigma_{sca}$ and $F_R$ spectra were examined by changing the NW diameter of the symmetric NW dimers and by varying the degree of asymmetry of the NW dimers.

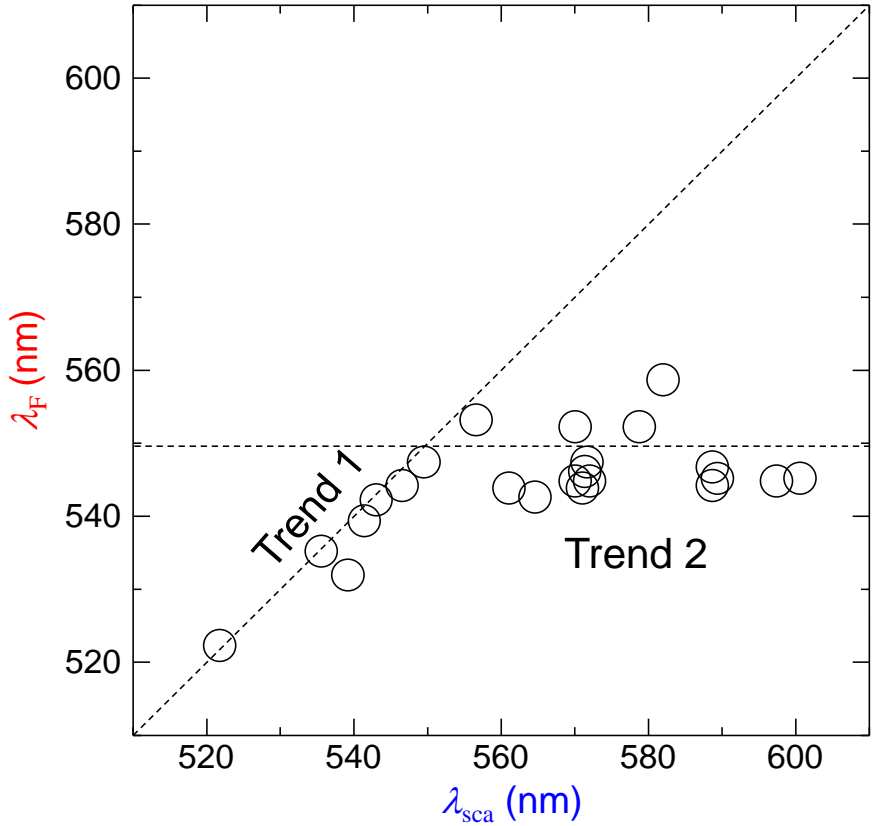

FIG. 8. Relationship between the $\lambda_{sca}$ and $\lambda_F$ for 23 single NW dimers exhibiting the two trends indicated by two black dashed lines.

Figures 9(a) and 9(b) show a setup of the FDTD calculation of a NW dimer with the coordinate system. The beam diameter for excitation light is infinite. The excitation polarization direction was set perpendicular to the long axis of the NW dimer because an intense $F_R$ of the 1D HS was generated along this direction by coupled plasmon resonance of both NWs.[23,24] The NW dimer comprised two cylindrical NWs with hemispherical ends of diameters $D_1$ and $D_2$. The gap distance was set to 0 nm, because the R6G molecules were added to NWs after their dimerization.[23] The lengths of the two NWs were 5.5 and 4.5 μm, respectively. Thus, the length of the 1D HS was 4.5 μm. The amplitude of the incident electric field $|E^I|$ is was set to be 1.0 V/m. Thus, the value of the calculated amplitude $|E^{loc}|$ was the same as the enhancement factor of electric field amplitude as $|E^{loc}/E^I|$. Note that the enhancement factor of electric field intensity $|E^{loc}/E^I|^2$ is equal to $F_R$ as shown in Eq. (1). Figure 9(c) exhibits $|E^{loc}|$ along y-axis at x, z = 0 nm. The two peaks in the $|E^{loc}|$ distribution at y = -9, 9 nm are set to be the y position of 1D HS. Figure 9(d) shows $|E^{loc}|$ and the phase retardation $\theta_{PR}$ of $E^{loc}$ against $E^I$ along the z-axis at x, y = 0, 9 nm. The value of y is set to be the position of 1D HS as in Fig. 9(c). The oscillation patterns in $|E^{loc}|$ and $\theta_{PR}$ of $E^{loc}$ are due to Fabry–Perot type interference.[23] Figure 9(e) shows the $\sigma_{sca}$ spectrum and $\theta_{PR}$ of the NW dimer. The value of $\theta_{PR}$ is the averaged from -2.0 to 2.0 μm along the z-axis of 1D HS. Figure 9(f) shows the $F_R$ spectrum and the $\theta_{PR}$ of the NW dimer. The value of $F_R$ is also averaged from -2.0 to 2.0 μm along the z-axis of the 1D HS. The spectra of $\sigma_{sca}$ and $F_R$ with the $\theta_{PR}$ were calculated along the 1D HSs for symmetric and asymmetric NW dimers to clarify the two trends observed in Fig. 8. First, these spectra of symmetric NW dimers are calculated by changing the diameter under the condition $D_1 = D_2$. Second, these spectra of asymmetric NW dimers are calculated by changing the ratios $D_1/D_2$ under the condition $D_1 = 60$ nm considering the distribution of NW diameters in Figs. 4(d) and 4(e). In these calculations, the nonlocal effect, which reduces $F_R$ by Landau damping owing to unscreened surface electrons,[36] was not considered because Landau damping does not change the spectral shape of $F_R$, but rather its intensity.[37]

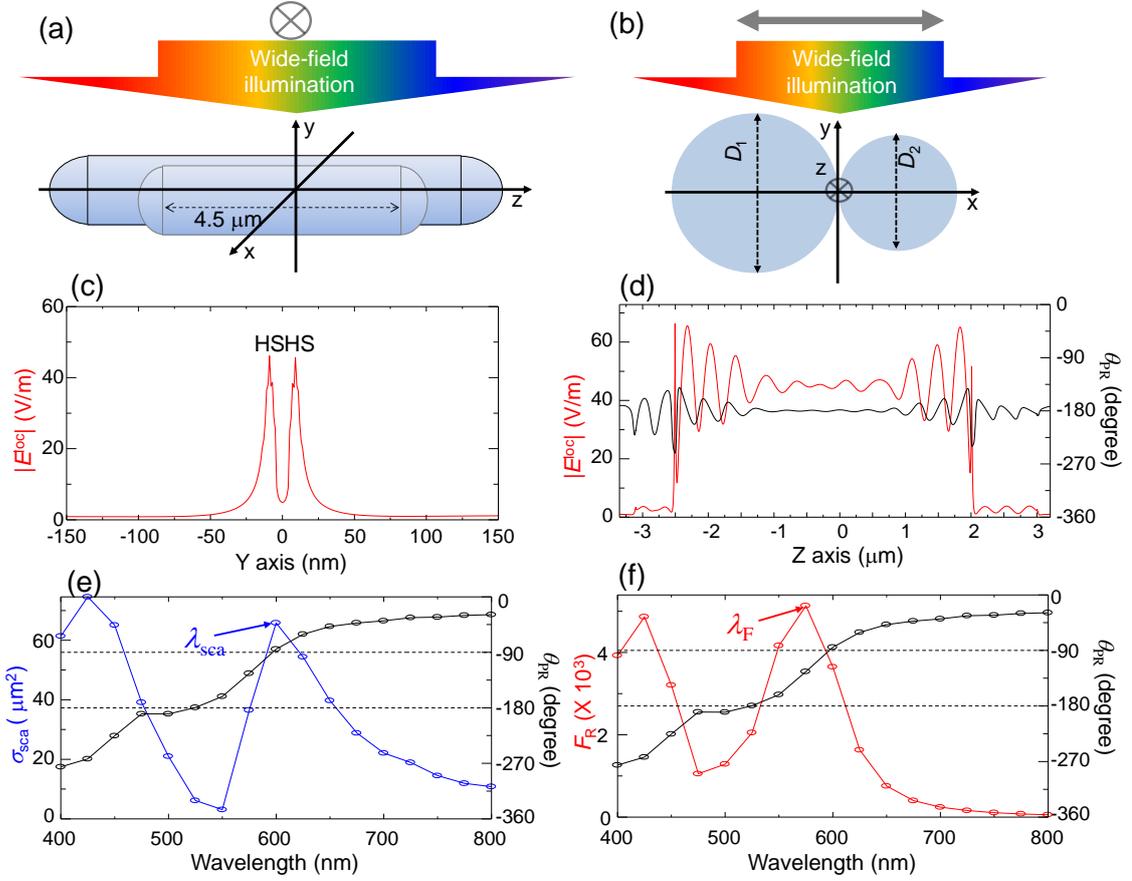

FIG. 9. (a) and (b) Image of an NW dimer with NW diameters $D_1$ and $D_2$, respectively, with length of 1D HS of 4.5 μm and direction of the excitation light beam (rainbow arrow) with the coordinate system used for the FDTD calculations. The gray circle with cross and gray arrow in (a) and (b) indicate the direction of excitation light polarization, respectively. The gap distance was set to 0 nm. (c) Distribution of $|E^{loc}|$ (red curve) along the y-axis at x, z = 0 and 0 nm calculated with $\lambda_{ex}$ of 525 nm. Two peaks, which are positions of HSs, are located at y = -9, 9 nm. (d) Distribution of $|E^{loc}|$ (red curve) with $\theta_{PR}$ (black curve) along the z axis at x = 0 and y = 9 nm, calculated with $\lambda_{ex}$ of 525 nm. (e) Calculated spectrum of $\sigma_{sca}$ (blue curve) with $\theta_{PR}$ (upper and lower dashed lines

indicate -90° and -180°, respectively) of 1D HS for the NW dimer. (f) Calculated spectrum of $F_R$ (red curve) with $\theta_{PR}$ (upper and lower dashed lines indicate -90° and -180°, respectively) of 1D HS for the NW dimer. The values of $F_R$ and $\theta_{PR}$ are averaged from -2.0 to 2.0 μm along the z-axis of 1D HS for (e) and (f). Positions of $\lambda_{sca}$ and $\lambda_F$ are indicated in (e) and (f), respectively.

Figures 10(a1)–10(a3) show the spectra of $\sigma_{sca}$ and $F_R$ of the symmetric NW dimers obtained by changing $D_1$ from 30 to 100 nm under the condition $D_1 = D_2$ with respect to the NW diameter distribution in Fig. 4(d). As $D_1$ increases, the positions of $\lambda_{sca}$ and $\lambda_F$ deviate from each other, and finally, $\lambda_F$ is located at the spectral dip of $\sigma_{sca}$. Figures 10(b1)–10(b3) show the spectra of $\sigma_{sca}$ with $\theta_{PR}$. The $\lambda_{sca}$ redshifts with increasing $D_1$, and is always located around the $\theta_{PR}$ of -90°, indicating that the maxima of $\sigma_{sca}$ corresponds to the superradiant dipole-dipole (DD) coupled plasmon resonance of the two NWs.[27-28] Figures 10(c1)–10(c3) show the spectra of $F_R$ and $\theta_{PR}$. The $\lambda_F$ redshifts with increasing $D_1$ and is located at a $\theta_{PR}$ of approximately -90° until 60 nm, indicating that $F_R$ is generated by DD-coupled plasmon resonance. However, the $\lambda_F$ does not shift even upon increasing $D_1$ to greater than 60 nm, and is finally located at a $\theta_{PR}$ of approximately -180°. The value of -180° and the deviation between $\lambda_{sca}$ and $\lambda_F$

indicate that $F_R$ for $D_1 > 60$ nm is generated by a subradiant plasmon by receiving light energy from DD-coupled plasmon through near-field interaction.[26-28] Furthermore, the same positions between $\lambda_F$ and the spectral dip of $\sigma_{sca}$ in Fig. 10(a3) also indicate that $F_R$ is generated by subradiant plasmon resonance.[26,27] Such subradiant resonance spectrally beside DD-coupled plasmon resonance has been attributed to dipole-quadrupole (DQ) coupled plasmon resonance.[27,38] The charge distribution of DD- and DQ-coupled plasmon is illustrated in Fig. 10(d). The spectral broadening in dipole resonance by increasing in NW diameters results in the spectral overlapping between dipole and quadrupole resonances and arises DQ coupled resonance.[27,38] Figure 10(e) shows the $D_1$ dependence of $\lambda_{sca}$ and $\lambda_F$ with $\theta_{PR}$ of -90° (DD-coupled plasmon resonance) and -180° (DQ-coupled plasmon resonance). The translucent red box indicates the range of the NW diameters: 40 nm < $D_1$ and $D_2$ < 80 nm by referring Fig. 4(d). Both $\lambda_{sca}$ and $\lambda_F$ follow the $\theta_{PR}$ of -90° for $D_1 < 60$ nm, indicating that the $F_R$ is generated by DD-coupled plasmon resonance. With $D_1 > 60$ nm, the $\lambda_{sca}$ keeps following $\theta_{PR}$ of -90°, but $\lambda_F$ moves to $\theta_{PR}$ of -180°, indicating that DQ-coupled plasmon resonance becomes dominant in generating $F_R$ for $D_1 > 60$ nm. Figure 10(f) shows the relationship between $\lambda_{sca}$ and $\lambda_F$ taking from Fig. 10(e) and the experimental relationship taking from Fig. 8. The translucent red box indicates the range of $\lambda_{sca}$ for 40 nm $< D_1$ and $D_2 <$ 80 nm. The two experimentally observed trends in Fig. 8 are clearly observed in the calculated relationship, indicating that the transition of the plasmon resonance generating $F_R$ from DD- to DQ-coupled resonance causes these two trends.

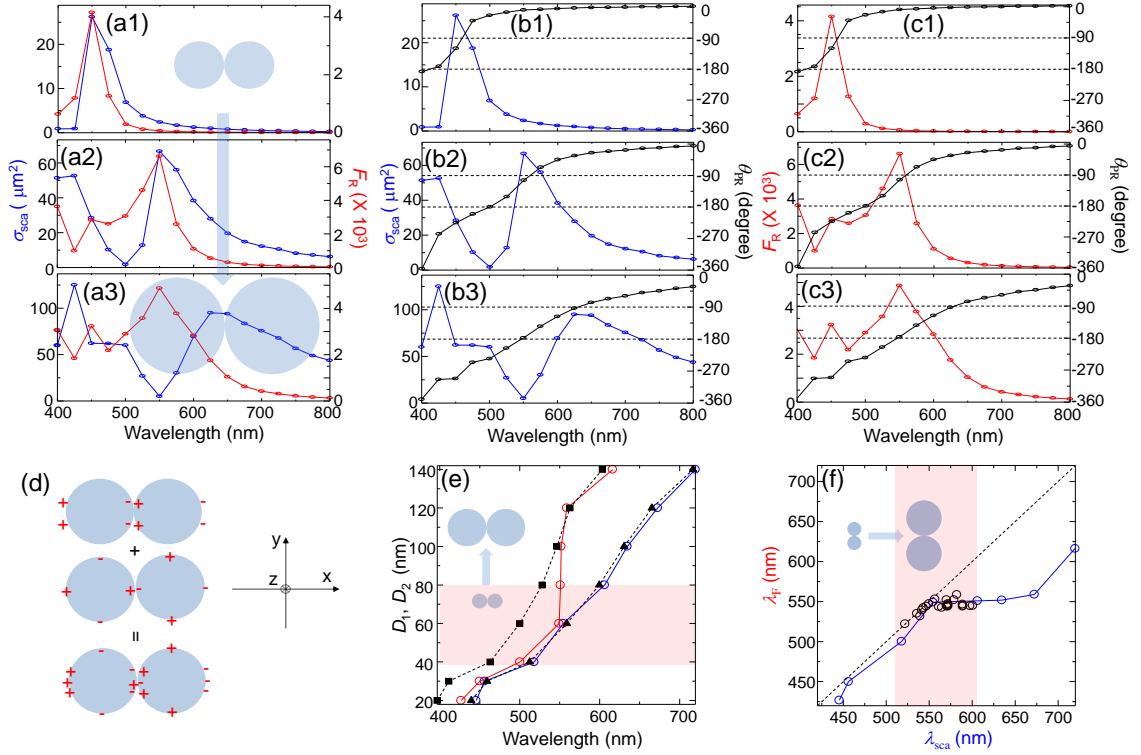

FIG. 10. (a1)–(a3) Diameter dependence of the spectra of $\sigma_{sca}$ (blue curves) and $F_R$ (blue curves) for NW dimers with $D_1$, $D_2$ of 30, 60, and 100 nm, respectively. (b1)–(b3) Diameter dependence the $\sigma_{sca}$ spectra (blue curves) with $\theta_{PR}$ (upper and lower dashed lines indicate -90° and -180, respectively) of 1D HS of NW dimers with $D_1$, $D_2$ of 30, 60, and 100 nm, respectively. (c1)–(c3) Diameter dependence of the $F_R$ spectra (red curves) with $\theta_{PR}$ (upper and lower dashed lines indicate -90° and -180°, respectively) of 1D HS of NW dimers with $D_1$, $D_2$ of 30, 60, and 100 nm, respectively. (d) Charge distributions

of dipole (upper), quadrupole (middle), and DQ-coupled (lower) resonance for symmetric NW dimer, respectively. The coordinate system is common to that shown in Fig. 9. (e) Diameter dependences of $\lambda_{sca}$ (blue open circles with line), $\lambda_F$ (red open circles with line), $\theta_{PR}$ = 90° (black close circles with dashed line, DD resonance), and $\theta_{PR}$ = 180° (black close triangles with dashed line, DQ resonance) for dimers with $D_1$, $D_2$ of 20 nm to 140 nm. The translucent red box indicates 40 nm < $D_1$, $D_2$ < 80 nm indicated from Fig. 4(d). (f) Calculated relationship between $\lambda_{sca}$ and $\lambda_F$ (blue open circles with line) for dimers with $D_1$, $D_2$ of 20 nm to 140 nm. Experimental relationship between $\lambda_{sca}$ and $\lambda_F$ (black open circles) taken from Fig. 8. The translucent red box indicates the range of $\lambda_{sca}$ expected by Figs. 4(d).

Figures 11(a1)–11(a3) show the spectra of $\sigma_{sca}$ and $F_R$ of the asymmetric NW dimers obtained by changing $D_1$ from 60 to 140 nm while maintaining $D_2$ = 60 nm with respect to the diameter ratios between the NW dimers in Fig. 4(e). As $D_1/D_2$ increased, the positions of $\lambda_{sca}$ and $\lambda_F$ deviated from each other, and both exhibited redshifts while maintaining their spectral deviation constant. Figures 11(b1)–11(b3) show the spectra of $\sigma_{sca}$ with $\theta_{PR}$. $\lambda_{sca}$ redshifts with increasing $D_1/D_2$ and is located at a $\theta_{PR}$ of -90°, indicating that the maxima of $\sigma_{sca}$ corresponds to the DD-coupled plasmon resonance of the two NWs.[27,28] Figures 11(c1)–11(c3) show the spectra of $F_R$ with $\theta_{PR}$. The $\lambda_F$ redshifts with increasing $D_1/D_2$ and is located at a $\theta_{PR}$ of approximately -90° until 60 nm, indicating that $F_R$ is generated by DD-coupled plasmon resonance.[27,28] However, the $\lambda_F$ redshifts at a $\theta_{PR}$ of -135°. The value of -135° in Figs. 11(c2) and 11(c3) and the constant deviation between $\lambda_{sca}$ and $\lambda_F$ in Figs. 11(a2) and 11(a3) indicate that the depolarization of DQ-coupled plasmon cannot be parallel to the incident light polarization. The possible charge distribution of such a situation is illustrated in Fig. 11(d).[39] As increasing diameter of the larger NW, its quadrupole resonance redshifts and is overlapped with dipole resonance of the smaller NW. The overlapping results in the DQ-coupled resonance.[27,28] We consider that the asymmetry in the charge distribution against the x-z plane results in $\lambda_F$ at a $\theta_{PR}$ of -135° not -180°. Figure 11(e) shows $D_1/D_2$ dependence of $\lambda_{sca}$ and $\lambda_F$ with $\theta_{PR}$ of -90°, -135°, and -180°. The translucent red box indicates the range 60 nm $< D_1 <$ 80 nm by considering Fig. 4(e). Both $\lambda_{sca}$ and $F_R$ follow the $\theta_{PR}$ of -90° at $D_1 = 60$ nm, indicating that the $F_R$ is generated by DD-coupled plasmon resonance. With increasing $D_1 > 60$ nm, $\lambda_{sca}$ keeps following $\theta_{PR}$ of -90°, but $\lambda_F$ moves to a $\theta_{PR}$ of -135°, indicating that the DQ-coupled plasmon as in Fig. 11(d) becomes dominant for generating $F_R$ for $D_1 > 60$ nm. Figure 11(f) shows the relationship between $\lambda_{sca}$ and $\lambda_F$ taking from Fig. 11(e), along with the experimental

relationship taking from Fig. 8. The translucent red box indicates the range of $\lambda_{sca}$ for 60 nm $< D_1 <$ 80 nm. The first trend was reasonably reproduced by the calculation. However, the second trend was not clearly observed in the calculated relationship, indicating that it was mainly caused by the diameter effect in Fig. 10 and not by the asymmetry effect in Fig. 11. This indication is supported by the distribution of the NW diameter, 40 nm $< D_1 <$ 90 nm, in Fig. 4(d) and the degree of asymmetry in Fig. 4(e) 0.5 $< D_1/D_2 <$ 2.0.

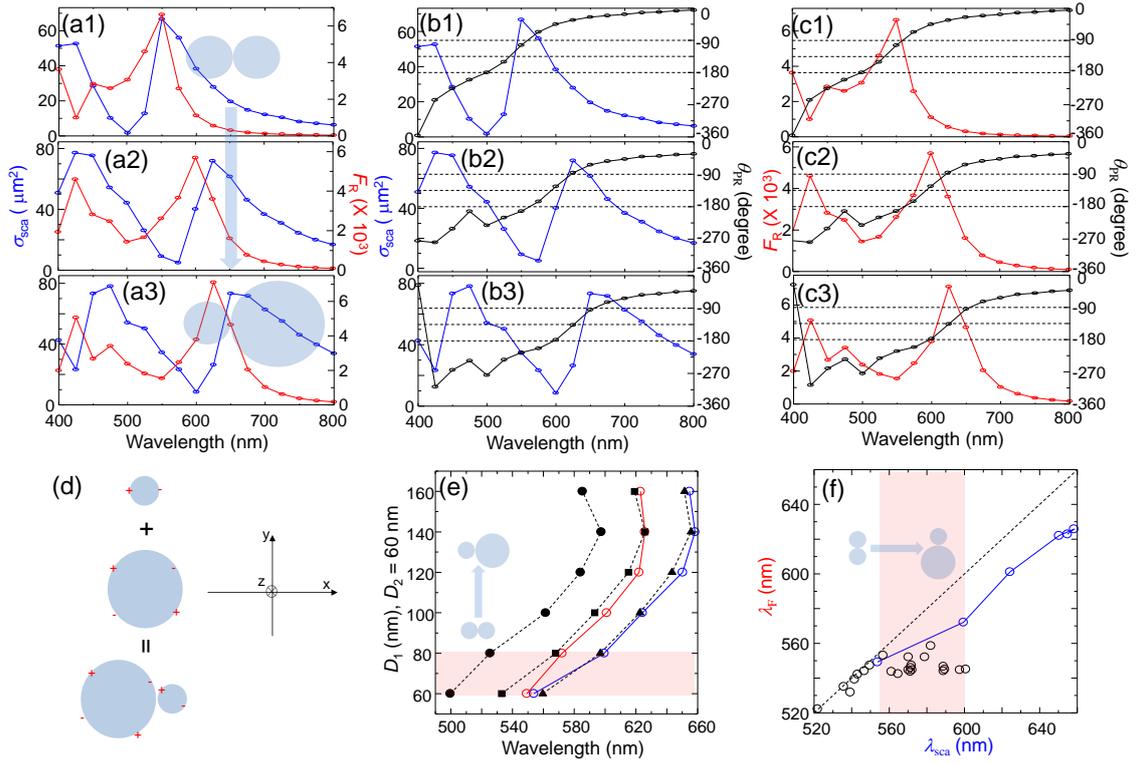

FIG. 11. (a1)–(a3) $D_1$ dependence of the spectra of $\sigma_{sca}$ (blue curves) and $F_R$ (red curves) for dimers with $D_1$ of 60, 100, and 140 nm, respectively, with maintaining $D_2 =$ 60 nm. (b1)–(b3) $D_1$ dependence of the $\sigma_{sca}$ spectra (blue curves) with $\theta_{PR}$ (upper,

middle, and lower dashed lines indicate -90°, -135° and -180°, respectively) of 1D HS of NW dimers with $D_1$ of 60, 100, and 140 nm, respectively, with $D_2 = 60$ nm. (c1)–(c3) $D_1$ dependence of the $F_R$ spectra (red curves) with $\theta_{PR}$ (upper, middle, and lower dashed lines indicate -90°, -135° and -180°, respectively) of 1D HS of NW dimers with $D_1$ of 60, 100, and 140 nm, respectively, with $D_2 = 60$ nm. (d) Charge distributions of dipole (upper), quadrupole (middle), and DQ-coupled (lower) resonance for asymmetric NW dimer, respectively. The coordinate system is common to that of Fig. 9. (e) $D_1$ dependences of $\lambda_{sca}$ (blue open circles with line), $\lambda_F$ (red open circles with line), $\theta_{PR} = 90°$ (black close circles with dashed line, DD resonance), $\theta_{PR} = 135°$ (black close squares with dashed line), and $\theta_{PR} = 180°$ (black close triangles with dashed line, DQ resonance) for dimers with $D_1$ of 60 nm to 140 nm, with maintaining $D_2$ of 60 nm. The translucent red box indicates the range of 60 nm $< D_1 <$ 80 nm indicated from Fig. 4(e). (f) Calculated relationship between $\lambda_{sca}$ and $\lambda_F$ (blue open circles with line) for dimers with $D_1$ of 60 nm to 140 nm with $D_2 = 60$ nm. Experimental relationship between $\lambda_{sca}$ and $\lambda_F$ (black open circles) taken from Fig. 8. The translucent red box indicates the range of $\lambda_{sca}$ expected by Fig. 4(e).

## V. Conclusion

We developed a spectroscopic technique to measure the $F_R$ of 1D HSs along single NW dimers using UFSEF of dye molecules by selecting $\lambda_{ex}$ suitable for observing whole spectral shapes of the $F_R$. The $F_R$ spectra were obtained by dividing the spectra of the UFSEF of single NW dimers by a reference UFSEF spectrum, in which the spectral variation in $F_R$ were averaged out. The derived $F_R$ spectra did not frequently match the $\sigma_{sca}$ spectra. This mismatch indicates that $F_R$ is not generated by superradiant plasmon resonance. These results were examined using FDTD calculations by changing the morphologies of symmetric and asymmetric NW dimers. The mismatch between the $F_R$ and $\sigma_{sca}$ spectra was reproduced by increasing the diameters of the symmetric NW dimers. The $\theta_{PR}$ analysis revealed that the DQ-coupled plasmon resonance, which is subradiant, mainly generates $F_R$ for dimers with NW diameters higher than 80 nm. This spectroscopic method, which directly measures $F_R$, is applicable to various plasmonic systems coupled with molecular excitons.[21,22,40] Particularly, this method will be useful for large plasmonic systems in which $F_R$ is dominated by subradiant modes.[41]

## Acknowledgements

This work was supported by a JSPS KAKENHI Grant-in-Aid for Scientific Research

(C) (grant number 21K04935). The authors thank to Prof. Jeyadevan Balachandran (the University of Shiga Prefecture) for silver nanowires.

**LIST OF ABBREVIATIONS**

$\hat{\mathbf{A}}$, Operators of vector potential

$D_1$, diameter of silver NW in Fig. 9(b)

$D_2$, diameter of silver NW in Fig. 9(b)

DD, dipole–dipole

DQ, dipole–quadrupole

$\Delta\omega_R$, radiative decay rates of plasmon resonance

$\Delta\omega_{NR}$, nonradiative decay rates of plasmon resonance

$E^I$, amplitudes of the incident electric field

$E^{loc}$, amplitudes of the enhanced local electric fields

EM, electromagnetic

$F$, Purcell factor

FDTD, finite-difference time-domain

$F_R$, electromagnetic enhancement factor in Eq. (1)

$\Gamma_{r0}$, total radiative decay rate of dye molecule in a free space

$\Gamma_{Rad}$, total radiative decay rates of a molecule enhanced by $F_R$

$\Gamma_{int}$, total internal relaxation rate of molecule

$\gamma_{r0}$, radiative decay rates at $\lambda_{em}$ of dye molecule in a free space

HS, hotspot

$n$, refractive index around the nanostructure

NA, numerical aperture

NP, nanoparticle

NW, nanowire

$\lambda$, wavelength of light

$\lambda_{em}$, wavelength of emission light

$\lambda_{ex}$, wavelength of excitation light

$\lambda_F$, wavelength of maximum of $F_R$

$\lambda_{sca}$, wavelength of maximum of $\sigma_{sca}$

1D HS, one-dimensional hotspot

$\hat{\mathbf{p}}$, operators of momentum

$Q$, quality factor of plasmon resonance

R6G, rhodamine 6G

SEA, surface-enhanced absorption

SEF, surface-enhanced fluorescence

SEM, scanning electron microscope

SE(R)RS, surface-enhanced (resonant) Raman scattering

$\sigma_{abs}$, absorption cross-section

$\sigma_{ext}$, extinction cross-section

$\sigma_{sca}$, Rayleigh scattering cross-section

$\sigma_{UFflo}$, cross-section of ultrafast fluorescence without EM enhancement

$\sigma_{UFSEF}$, cross-section of UFSEF

$\theta_{PR}$, phase retardation of $E^{loc}$ against $E^I$

UFSEF, ultrafast surface enhanced fluorescence

$V_P$, mode volume of plasmon

**References**

[1] T. Itoh, Y. S. Yamamoto, and Y. Ozaki, Plasmon-enhanced spectroscopy of absorption and spontaneous emissions explained using cavity quantum optics, Chem. Soc. Rev. **46**, 3904 (2017).

[2] H. Xu, E. J. Bjerneld, M. Käll, and L. Börjesson, Spectroscopy of single haemoglobin molecules by surface enhanced Raman scattering, Phys. Rev. Lett. **83**, 4357 (1999).

[3] W. Zhu and K. B. Crozier, Quantum mechanical limit to plasmonic enhancement as observed by surface-enhanced Raman scattering. Nat. Commun. **5**, 5228 (2014).

[4] C. Ciraci, R. T. Hill, J. J. Mock, Y. Urzhumov, A. I. Fernandez-Dominguez, S. A. Maier, J. B. Pendry, A. Chilkoti, D. R. Smith, Probing the ultimate limits of plasmonic enhancement, Science **337**, 1072–1074 (2012).

[5] S. Nie and S. Emory, Probing single molecules and single nanoparticles by surface-enhanced Raman scattering, Science **275**, 1102 (1997).

[6] K. Kneipp, Y. Wang, H. Kneipp, L. Perelman, I. Itzkan, R. R. Dasari, and M. Feld, Single molecule detection using surface-enhanced Raman scattering (SERS), Phys. Rev. Lett. **78**, 1667 (1997).

[7] A. M. Michaels, M. Nirmal, and L. E. Brus, Surface enhanced Raman spectroscopy of individual rhodamine 6G molecules on large Ag nanocrystals, J. Am. Chem. Soc. **121**, 9932 (1999).

[8] E. C. Le Ru, M. Meyer, and P. G. Etchegoin, Proof of single-molecule sensitivity in surface enhanced Raman scattering (SERS) by means of a two-analyte technique, J. Phys. Chem. B **110**, 1944 (2006).

[9] A. B. Zrimsek, A. I. Henry, and R. P. Van Duyne, Single molecule surface-enhanced Raman spectroscopy without nanogaps, J. Phys. Chem. Lett. **4**, 3206 (2013).

[10] J. J. Baumberg, J. Aizpurua, M. H. Mikkelsen, and D. R. Smith, Extreme nanophotonics from ultrathin metallic gaps, Nat. Mater. **18**, 668 (2019).

[11] T. Itoh, Y. S. Yamamoto, H. Tamaru, V. Biju, S. Wakida, and Y. Ozaki, Single-molecular surface-enhanced resonance Raman scattering as a quantitative probe of local electromagnetic field: The case of strong coupling between plasmonic and excitonic resonance, Phys. Rev. B **89**, 195436 (2014).

[12] T. Itoh and Y. S. Yamamoto, Reproduction of surface-enhanced resonant Raman scattering, and fluorescence spectra of a strong coupling system composed of a single silver nanoparticle dimer and a few dye molecules, J. Chem. Phys. **149**, 244701 (2018).

[13] T. Itoh, Y. S. Yamamoto, and T. Okamoto, Anti-crossing property of strong coupling

system of silver nanoparticle dimers coated with thin dye molecular films analyzed by electromagnetism, J. Chem. Phys. **152**, 054710 (2020).

[14] T. Itoh and Y. S. Yamamoto, Between plasmonics and surface-enhanced resonant Raman spectroscopy: toward single-molecule strong coupling at a hotspot, Nanoscale **13**, 1566 (2021).

[15] R. Esteban, J. J. Baumberg, and J. Aizpurua, Molecular optomechanics approach to surface-enhanced Raman scattering, Acc. Chem. Res. **55**, 1889–1899 (2022).

[16] R. F. Ribeiro, L. A. Martínez-Martínez, M. Du, J. Campos-Gonzalez-Angulo, and J. Yuen-Zhou, Polariton chemistry: controlling molecular dynamics with optical cavities. Chem. Sci. **9**, 6325 (2018).

[17] K. Yoshida, T. Itoh, H. Tamaru, V. Biju, M. Ishikawa, and Y. Ozaki, Quantitative evaluation of electromagnetic enhancement in surface-enhanced resonance Raman scattering from plasmonic properties and morphologies of individual Ag nanostructures, Phys. Rev. B **81**, 115406 (2010).

[18] K. J. Savage, M. M. Hawkeye, R. Esteban, A. G. Borisov, J. Aizpurua, and J. J. Baumberg, Revealing the quantum regime in tunnelling plasmonics, Nature **491**, 574 (2012).

[19] T. Itoh, M. Iga, H. Tamaru, K. Yoshida, V. Biju, and M. Ishikawa, Quantitative

evaluation of blinking in surface enhanced resonance Raman scattering and fluorescence by electromagnetic mechanism, J. Chem. Phys. **136**, 024703 (2012).

[20] T. Itoh, Y. S Yamamoto, V. Biju, H. Tamaru, and S. Wakida, Fluctuating single $sp^2$ carbon clusters at single hotspots of silver nanoparticle dimers investigated by surface-enhanced resonance Raman scattering, AIP Adv. **5**, 127113 (2015).

[21] S. Y. Ding, E. M. You, Z. Q. Tian, and M. Moskovits, Electromagnetic theories of surface-enhanced Raman spectroscopy, Chem. Soc. Rev. **46**, 4042–4076 (2017).

[22] T. Itoh, M. Prochazka, Z.-C. Dong, W. Ji, Y. S. Yamamoto, Y. Zhang, and Y. Ozaki, Toward a new era of SERS and TERS at the nanometer scale: from fundamentals to innovative applications, Chem. Rev. **123**, 1552 (2023).

[23] T. Itoh, T. Y. S. Yamamoto, Y. Kitahama, and J. Balachandran, One-dimensional plasmonic hotspots located between silver nanowire dimers evaluated by surface-enhanced resonance Raman scattering, Phys. Rev. B **95**, 115441 (2017).

[24] T. Itoh, T. Y. S. Yamamoto, and J. Balachandran, Propagation mechanism of surface plasmons coupled with surface-enhanced resonant Raman scattering light through a one-dimensional hotspot along a silver nanowire dimer junction, Phys. Rev. B **103**, 245425 (2021).

[25] J. J. Sakurai, *Advanced Quantum Mechanics* (Addison-Wesley, Reading, MA, 1967);

W. Heitler, *The Quantum Theory of Radiation* (Oxford Univ. Press, MA, 1954).

[26] T. Itoh, Y. S. Yamamoto, Demonstration of electromagnetic enhancement correlated to optical absorption of single plasmonic system coupled with molecular excitons using ultrafast surface-enhanced fluorescence, J. Chem. Phys. **159**, 2 034709 (2023).

[27] T. Itoh and Y. S. Yamamoto, Contribution of subradiant plasmon resonance to electromagnetic enhancement in resonant Raman with fluorescence examined by single silver nanoparticle dimers, J. Phys. Chem. C **127**, 5886–5897 (2023)

[28] T. Itoh and Y. S. Yamamoto, Correlated polarization dependences between surface-enhanced resonant Raman scattering and plasmon resonance elastic scattering showing spectral uncorrelation to each other, J. Phys. Chem. B **127**, 4666–4675 (2023).

[29] T. Itoh, Y. S. Yamamoto, and T. Okamoto, Absorption cross-section spectroscopy of a single strong-coupling system between plasmon and molecular exciton resonance using a single silver nanoparticle dimer generating surface-enhanced resonant Raman scattering, Phys. Rev. B, **99**, 235409 (2019).

[30] C. F. Bohren, and D. R. Huffman, *Absorption and Scattering of Light by Small Particles* (Wiley, New York, 1983).

[31] E. C. Le Ru, P. G. Etchegoin, J. Grand, N. Félidj, J. Aubard, and G. Lévi, Mechanisms of spectral profile modification in surface-enhanced fluorescence, J. Phys. Chem. C **111**,

44, 16076–16079 (2007).

[32] T. Itoh, Y. S. Yamamoto, H. Tamaru, V. Biju, N. Murase, and Y. Ozaki, Excitation laser energy dependence of surface-enhanced fluorescence showing plasmon-induced ultrafast electronic dynamics in dye molecules, Phys. Rev. B **87**, 235408 (2013).

[33] B. Li, S. Ye, I. E. Stewart, S. Alvarez, and B. J. Wiley, Synthesis and Purification of Silver Nanowires To Make Conducting Films with a Transmittance of 99%, Nano Lett. **15**, 6722 (2015).

[34] P. Lee and D. Meisel, Adsorption and surface-enhanced Raman of dyes on silver and gold sols, J. Phys. Chem. **86**, 3391 (1982).

[35] P. B. Johnson, and R. W. Christy, Optical constants of the noble metals, Phys. Rev. B **6**, 4370–4379 (1972).

[36] P. Johansson, H. Xu, and M. Käll, Surface-enhanced Raman scattering and fluorescence near metal nanoparticles, Phys. Rev. B **72**, 035427 (2005).

[37] C. M. Galloway, P. G. Etchegoin, and E. C. Le Ru, Phys. Rev. Lett. 103, 063003 (2009).

[38] Y. Tanaka, A. Sanada, K. Sasaki, Nanoscale interference patterns of gap-mode multipolar plasmonic fields, Sci. Rep. **2**, 764 (2012).

[39] L. V. Brown, H. Sobhani, J. B. Lassiter, P. Nordlander, N. J. Halas, Heterodimers:

plasmonic properties of mismatched nanoparticle pairs, ACS Nano **4**, 819–832 (2010).

[40] J. Langer et al., Present and future of surface-enhanced Raman scattering, ACS Nano **14**, 28–117 (2020).

[41] S. Zhang, K. Bao, N. J. Halas, H. Xu, and P. Nordlander, Substrate-Induced Fano resonances of a plasmonic nanocube: a route to increased-sensitivity localized surface plasmon resonance sensors revealed, Nano Lett. **11**, 1657–1663 (2011); M. Pellarin, J. Ramade, J. M. Rye, C. Bonnet, M. Broyer, M.-A. Lebeault, J. Lermé, S. Marguet, J. R.G. Navarro, and E. Cottancin, Fano transparency in rounded nanocube dimers induced by gap plasmon coupling, ACS Nano **10**, 11266−11279 (2016).